\documentclass[aps,pra,showpacs,twocolumn]{revtex4}

\usepackage{amsmath, amssymb}
\usepackage{color, graphicx}
\usepackage{amsmath, amssymb}

\begin{document}
\title{Entanglement quasiprobabilities of squeezed light}

\author{J. Sperling}
\author{W. Vogel}
\affiliation{Arbeitsgruppe Quantenoptik, Institut f\"ur Physik, Universit\"at Rostock, D-18051 Rostock, Germany}


\begin{abstract}
We demonstrate the feasibility to completely characterize entanglement by negativities of quasiprobabilities.
This requires the complete solution of a sophisticated mathematical problem, the so-called separability eigenvalue problem.
Its solution is obtained for a non-Gaussian continuous variable quantum state, a two-mode squeezed state undergoing dephasing. 
This is a standard scenario for experiments with quantum-correlated radiation fields.
\end{abstract}

\maketitle

\section{Introduction}\label{Sec:intro}
Entanglement is a key resource of quantum technology, for an introduction see~\cite{Horodecki,cite6,Chuang}.
The nonclassical properties of entangled quantum states have been discussed since the early days of quantum physics~\cite{EPR,Schroedinger}.
Applications of entangled states are, for example, quantum key distribution~\cite{cite3}, quantum dense coding~\cite{cite4}, and quantum teleportation~\cite{cite5}.

The most prominent and frequently used continuous variable (CV) entangled quantum state is the two-mode squeezed-vacuum state.
It has been considered, for example, as a resource for quantum teleportation~\cite{cite21,cite20}, quantum dense coding~\cite{cite22}, and quantum memories~\cite{cite23}.
It has been shown that this state violates a continuous variable Bell inequality~\cite{cite24}.
Noise and loss effects of pure two-mode squeezed-vacuum states have been studied, such as noise due to the transmission in optical fibers~\cite{cite26}, or dephasing and amplitude damping~\cite{cite25}.

In general, nonclassical effects in Quantum Optics are characterized by the most prominent example of a quasiprobability distribution, the Glauber-Sudarshan $P$~function~\cite{GS2,GS1}.
Negativities of this distribution verify that the corresponding quantum state cannot be interpreted as a classical mixture of coherent states, the latter being the closest analog to the classical behavior.
Sometimes the negativities are hidden in the highly singular structure of the $P$~function.
Recently, it has been shown in theory~\cite{thomas1} and in experiment~\cite{thomas2} that these negativities can be revealed by filtered, regularized quasiprobabilities.

The problem to verify the property of entanglement of a quantum state is very complex in general.
Most difficult is the situation for non-Gaussian CV entangled states. One possibility is the use of matrices of higher-order moments~\cite{Sh-Vo}, which has been successfully applied recently~\cite{Walborn}.
Alternatively, entanglement witnesses can identify any kind of entanglement~\cite{Peres1,Spe2}. 
However, the first method needs to identify particular matrices of moments out of an infinite manifold. In the second method one has to find the optimal witness among an uncountable number of Hermitian operators.
Both types of entanglement tests are very difficult for non-Gaussian states. 

In the present article, we aim to provide methods of quasiprobability distributions for the property entanglement. Explicit solutions are derived for phase-diffused two-mode squeezed vacuum states, which are nontrivial examples of non-Gaussian CV entangled states.
Our approach is based upon the previously developed concept of entanglement quasiprobabilities which requires to
solve the so-called separability eigenvalue (SE) equations for the quantum state under study~\cite{Spe1}.
Negativities of these quasiprobabilities for a given state have been shown to be equivalent to entanglement.
The general solution of this problem will be derived for our example of a class of non-Gaussian CV entangled states, which is of great interest for applications in quantum technology. Our method solely requires the quantum state and the solution of its SE problem, avoiding the complex identification of moments or witness operators.
Note that the connection between quasiprobabilities and their applications to quantum technology has been reviewed in Ref.~\cite{QuasiReview}.

In Sec.~\ref{Sec:ExpSetup}, we consider a realistic scenario for the generation of the considered kinds of states.
We recapitulate the basic reconstruction procedure for the entanglement quasiprobability in Sec.~\ref{Sec:EntQP}.
The general solution of the separability eigenvalue equations is derived in Sec.~\ref{Sec:SolSEE}.
On this basis we study in Sec.~\ref{Sec:Visu} the resulting entanglement quasiprobabilities, which visualize entanglement via their negativities.
A brief summary and conclusions are given in Sec.~\ref{Sec:SandC}.

\section{Experimental situation}\label{Sec:ExpSetup}
In the following we consider the generation and the characterization of a nontrivial class of bipartite entangled CV states.
They are created via phase randomization of Gaussian two-mode squeezed-vacuum states, which eventually leads to so-called non-Gaussian states. 
They are not pure states anymore, but an infinite mixture of pure states.
Altogether the states have complex properties, so that an entanglement test is a demanding task.
In the following we will not only witness the entanglement, but we visualize the entanglement by negativities of entanglement quasiprobabilities.
The latter are necessary and sufficient for entanglement and contain the complete information on the quantum state.

\begin{center}
	\begin{figure}[ht]
		\includegraphics[width=5cm]{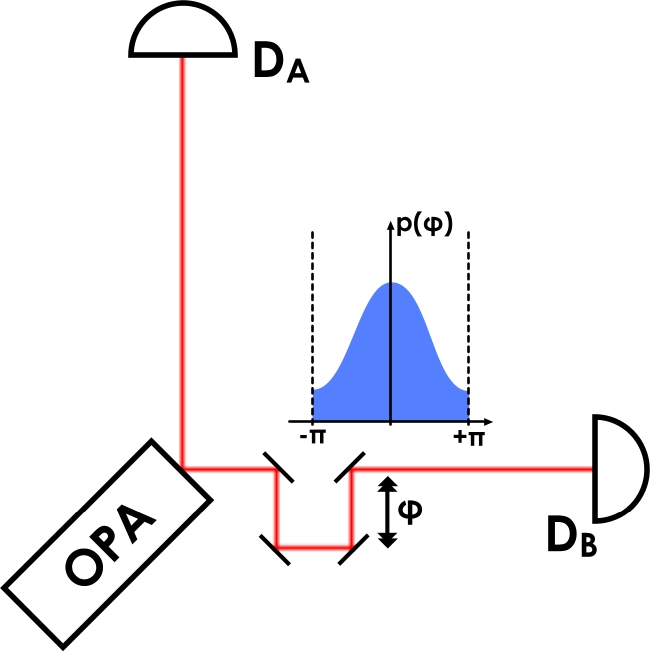}
		\caption{A two-mode squeezed-vacuum state is generated by an optical parametric amplifier (OPA).
		One mode is phase randomized by a random change of the optical path length.
		Correlation measurements are performed in both modes (${\rm D_A}$ and ${\rm D_B}$) for a reconstruction of the density matrix in Fock basis.}\label{Fig:Setup}
	\end{figure}
\end{center}

Let us consider the generation of a two-mode squeezed-vacuum by an optical parametric amplifier~\cite{Schnabel1},
\begin{align}
	\varrho_{0}=(1-\zeta^2)\sum_{m,n\in\mathbb N}\zeta^{m+n}|m,m\rangle\langle n,n|,
\end{align}
which is a pure entangled state and $\zeta$ is the squeezing parameter ($0<\zeta<1$).
Let us consider an additional phase randomization in one of the channels~\cite{Schnabel2}, cf. the scheme in Fig.~\ref{Fig:Setup}.
We may model the phase randomization by a local unitary phase transformation $\mathbb I_A\otimes U(\varphi)$ which is performed with a certain probability $p_\sigma(\varphi)$. As an example, we will consider a $2\pi$-periodic Gaussian phase distribution with variance $\sigma$,
\begin{align}
	U(\varphi)=&\sum_{m\in\mathbb N} \exp\left[{\rm i}m\varphi\right]|m\rangle\langle m|,\\
	p_\sigma(\varphi)=&\sum_{p\in\mathbb Z} \frac{\exp\left[-\frac{(\varphi+2\pi p)^2}{2\sigma^2}\right]}{\sqrt{2\pi\sigma^2}}.\label{Eq:PeriodicGaussian}
\end{align}
Due to the central limit theorem, such a distribution may often occur in practice, whenever  small dephasing effects are added up.
The resulting phase randomization probability is a Gaussian distribution with respect to the phase~$\varphi$.
In phase space (position-momentum representation), however, $p_\sigma(\varphi)$ is not a Gaussian function anymore.

The dephased squeezed state is given by
\begin{align}
	\nonumber \varrho_\sigma=&\int_0^{2\pi} d\varphi\,p_\sigma(\varphi)\,(\mathbb I_A\otimes U(\varphi))\varrho_{0}(\mathbb I_A\otimes U(\varphi)^\dagger)\\
	\label{eq:dephased-sv} =&\sum_{m,n\in\mathbb N} (1-\zeta^2)\zeta^{m+n}\\
	\nonumber&\quad\quad\quad\times\exp\left[-\frac{\sigma^2(m-n)^2}{2}\right]\,|m,m\rangle\langle n,n|.
\end{align}
Such a state $\varrho_{\sigma}$ is, in general, experimentally accessible, cf.~\cite{Schnabel1,Schnabel2}.
For related states, the quantification of entanglement has been investigated theoretically~\cite{Spe3,Spe4}, and entanglement tests have been experimentally performed in reduced subspaces $2\otimes2$, for detail see~\cite{Schnabel3}.
However, in a $2\otimes2$ Hilbert space the Peres-Horodecki entanglement test is necessary and sufficient~\cite{Peres1,Peres2}.
In this paper, we go beyond this limitation.
Moreover, the entanglement test is given in a form that includes the full information on the quantum state.

\section{Determination of entanglement quasiprobabilities}\label{Sec:EntQP}
A first indication of the existence of an entanglement quasiprobability was the proof of the following representation of general entangled states $\rho$ in terms of separable ones:
\begin{align}
	\rho=(1+\lambda)\rho_{\rm sep}-\lambda\rho_{\rm sep}',
\end{align}
with $\lambda>0$, $\rho_{\rm sep}$ and $\rho_{\rm sep}'$ being separable states~\cite{entQPpre1,entQPpre2}.
Recently, we have shown how to construct entanglement quasiprobabilities on this basis~\cite{Spe1}.
This method enables us to show that an entangled quantum state $\rho$ can be written in terms of separable states $|a,b\rangle\langle a,b|$ and a particular quasiprobability $P_{\rm Ent}(a,b)$,
\begin{align}
	\rho=\int dP_{\rm Ent}(a,b)|a,b\rangle\langle a,b|,\label{eq:ent-repr}
\end{align}
in analogy to the Glauber-Sudarshan $P$~function.
It is worth to mention that we can always rewrite the integral as an infinite (sometimes finite) sum,
\begin{align}\label{Eq:DefDecPEnt}
	\rho=\sum_k P_{\rm Ent}(a_k,b_k)|a_k,b_k\rangle\langle a_k,b_k|,
\end{align}
with some $P_{\rm Ent}(a_k,b_k)<0$ for an entangled state $\rho$.

The initial difficulty in defining such an entanglement quasiprobability $P_{\rm Ent}$ was the mathematical problem that the decomposition of general states in terms separable ones is not unique.
This means that a non-optimized decomposition of a separable state may contain negativities.
Our method overcomes the ambiguity of the decomposition by optimization, based on the solution of the separability eigenvalue problem of the quantum state.
The state is entangled, if and only if the optimized $P_{\rm Ent}$ contains negativities.

\subsection{Separability eigenvalue equations.}
Before we consider the reconstruction of $P_{\rm Ent}$, let us recall a mathematical tool which is needed to obtain the optimized entanglement quasiprobability.
This tool is the set of separability eigenvalue (SE) equations~\cite{Spe1,Spe2},
\begin{align}\label{Eq:SEE}
	\rho_b|a\rangle=g|a\rangle \quad\text{and}\quad \rho_a|b\rangle=g|b\rangle,
\end{align}
with the reduced operators $\rho_a={\rm Tr}_A([|a\rangle\langle a|\otimes \mathbb I_B]\rho)$ and $\rho_b={\rm Tr}_B([\mathbb I_A\otimes|b\rangle\langle b|]\rho)$, with $\mathbb I_{A,B}$ being the unity operator in subsystem $A$,$B$.
Using these equations one can, for example, construct all bipartite entanglement witnesses~\cite{Spe1}.
Applying the SE equations to a rank one operator~\cite{Spe2}, e.g. $|\psi\rangle\langle\psi|$,  yields the well-known Schmidt decomposition of $|\psi\rangle$~\cite{Chuang}.
The value $g$ in Eq.~(\ref{Eq:SEE}) is referred to as separability eigenvalue (SE value), and the vector $|a,b\rangle$ denotes the separability eigenvector (SE vector).

The notion and properties of the SE equations and its solutions are in full analogy to the well-known {\em ordinary} eigenvalue equations.
Solving the {\em ordinary} eigenvalue problem, $\rho|\psi_k\rangle=p_k|\psi_k\rangle$, enables us to find the spectral decomposition of a quantum state $\rho$ in terms of pure states $|\psi_k\rangle\langle\psi_k|$ and non-negative eigenvalues $p_k$.
Whenever an eigenvalue is negative, the corresponding operator cannot be a valid density operator.
In the present case we obtain the optimal representation of $\rho$ in terms of product states defined in Eq.~(\ref{eq:ent-repr}).
Whenever the weight factor $P_{\rm Ent}$ contains negativities, the state cannot be separable; hence it is entangled.

\subsection{Reconstruction of $P_{\rm Ent}$.}
As a fundamental application of the SE equations, let us study a three step reconstruction scheme to obtain $P_{\rm Ent}$, cf.~\cite{Spe1}.
In a first step, we solve the SE equation for the quantum state $\rho$,
\begin{enumerate}
	\item[(i)] $\rho_{b_k}|a_k\rangle=g_k|a_k\rangle$ and $\rho_{a_k}|b_k\rangle=g_k|b_k\rangle$.
\end{enumerate}
Note that all SE values $g_k$ are non-negative, and a direct decomposition of $\rho$ as in the case of the spectral decomposition is impossible.
In a second and third step we can obtain the entanglement quasiprobability $P_{\rm Ent}$.
We define 
\begin{enumerate}
	\item[(ii)] the Matrix ${\bf G}=(|\langle a_k,b_k|a_l,b_l\rangle|^2)_{k,l}$,\\
	a vector of the SE values $\vec g=(g_k)_k$,\\
	and a quasiprobability vector $\vec p=(P_{\rm Ent}(a_l,b_l))_l$.
\end{enumerate}
Using these definitions, we simply invert the linear equation
\begin{enumerate}
	\item[(iii)] ${\bf G}\vec p=\vec g$
\end{enumerate}
to obtain $P_{\rm Ent}$.
Finally, the quantum state under study can be written as given in Eq.~(\ref{Eq:DefDecPEnt}).

Performing the steps (i) -- (iii), we obtain the optimized $P_{\rm Ent}$ of a given state $\rho$.
At least one negative value in the determined $P_{\rm Ent}$ is necessary and sufficient to verify that the corresponding quantum state is entangled.
In analogy to the spectral decomposition, the quantum state under study is not a valid separable state.

\section{Solution of SE equations}\label{Sec:SolSEE}
For general quantum states the solution of the SE equations is a cumbersome mathematical problem, for which no systematic methods are available yet.
In this section, we solve this problem for a class of density operators of the form
\begin{align}
	\rho=\sum_{m,n\in\mathbb N} \rho_{m,n} |m,m\rangle\langle n,n|,
\label{eq:consideredstates}
\end{align}
with $\rho_{m,n}=\rho_{n,m}^\ast$. To this class belong the  
two-mode squeezed-vacuum states and their dephased versions. The solution is a substantial 
progress, since among the practically used CV quantum states the squeezed ones play a dominant role.
The characterization of dephasing includes important effects which typically occur during the propagation of the radiation field.

\subsection{General Solution}  
We seek the  general solution of the SE equations for quantum states of the type given in Eq.~(\ref{eq:consideredstates}).
The reduced operators for the states 
\begin{align}
|a\rangle=\sum_k a_k|k\rangle \quad\text{and}\quad |b\rangle=\sum_l b_l |l\rangle 
\end{align}
can be written as
\begin{align}
	\rho_a=&{\rm Tr}_A([|a\rangle\langle a|\otimes \mathbb I_B)]\rho)=\sum_{m,n} a_m^\ast\rho_{m,n} a_n|m\rangle\langle n|,\\
	\rho_b=&{\rm Tr}_B([\mathbb I_A\otimes|b\rangle\langle b|)]\rho)=\sum_{m,n} b_m^\ast\rho_{m,n} b_n|m\rangle\langle n|.
\end{align}
The formulation of the SE equations~(\ref{Eq:SEE}) in Fock basis yields, for $\rho_a|b\rangle=g|b\rangle$:
\begin{align}
	\label{Eq:SEModB} 
	\sum_m a^\ast_m \left[\sum_n \rho_{m,n} {a_nb_n}\right] |m\rangle=\sum_m gb_m|m\rangle,
\end{align}
and for $\rho_b|a\rangle=g|a\rangle$:
\begin{align}
	\label{Eq:SEModA} 
	\sum_m b^\ast_m \left[\sum_n \rho_{m,n} {a_nb_n}\right] |m\rangle=\sum_m ga_m|m\rangle.
\end{align}
Equivalently, for all $m\in\mathbb N$ holds
\begin{align}
	\left[\sum_{n\in\mathbb N} \rho_{m,n} a_nb_n\right]a_m^\ast=&g\,b_m,\\
	\left[\sum_{n\in\mathbb N} \rho_{m,n} a_nb_n\right]b_m^\ast=&g\,a_m,
\end{align}
with normalizations $\sum_{n\in\mathbb N} |a_n|^2=\sum_{n\in\mathbb N} |b_n|^2=1$.

This system of equations can be solved by inserting the equations into each other.
We obtain
\begin{align}
	\left(\left|\sum_{n\in\mathbb N}\rho_{m,n}a_nb_n\right|^2-{g^2}\right)a_m=0,
\end{align}
and analogous relations for $b_m$.
Now, the solution can be given.
For all $m\in\mathbb N$ holds
\begin{align}\label{Eq:CondZero}
	\left|\sum_{n\in\mathbb N}\rho_{m,n}a_nb_n\right|=g\quad
	\text{or}\quad a_m=b_m=0.
\end{align}
which already delivers all solutions.
This given form of the solution can be interpreted as a relation between different coefficients of the solution.
Either the projections $|\langle m,m|\rho|a,b\rangle|=g$ are equal for different values of $m$, or the component does not contribute to $g$ at all, for $a_m,b_m=0$.

In the following, we rewrite the general solution in a more convenient form for practical calculations.
For this purpose we introduce the set $\mathcal N$ which includes all indices $m$ fulfilling the second part of condition~(\ref{Eq:CondZero}),
\begin{align}
	\mathcal N=\{m\in\mathbb N: a_m=b_m=0\}.
\end{align}
The remaining indices are elements of the set $\overline{\mathcal N}=\mathbb N\setminus\mathcal N$.
For those elements the first part of condition~(\ref{Eq:CondZero}) can be written as
\begin{align}\label{Eq:SolutionSetN}
	\sum_{n\in\overline{\mathcal N}} \rho_{m,n} a_n b_n=g\,e^{{\rm i}\phi_m},
\end{align}
for arbitrary phases $0\leq\phi_m<2\pi$.
In addition, we introduce some abbreviation for two vectors $c$ and $e$, and the coefficient matrix $\rho_{\overline{\mathcal N}}$, 
\begin{align}
	\nonumber c=(a_nb_n)_{n\in\overline{\mathcal N}},\\
	e=(e^{{\rm i}\phi_m})_{m\in\overline{\mathcal N}},\\
	\nonumber \rho_{\overline{\mathcal N}}=(\rho_{m,n})_{m,n\in\overline{\mathcal N}}.
\end{align}
Note that we have changed the normalization of the separable states, $a_mb_m\to a_mb_m/g$.
This enables us to get rid of $g$.
A trivial normalization of the unormalized vectors at the end of the procedure delivers the correct states.
We simply rewrite Eq.~(\ref{Eq:SolutionSetN}) as
\begin{align}
	\rho_{\overline{\mathcal N}}c= e
	\quad\Leftrightarrow\quad
	c=\rho_{\overline{\mathcal N}}^{-1}e.
\end{align}
We obtain our SE vectors from the relation $c_m=a_mb_m$, including the normalization. The SE values follow from the fact that $g=\langle a,b|\rho|a,b\rangle$, cf. Eq.~(\ref{Eq:SEE}).
We can choose now arbitrary subsets $\mathcal N$ of all non-negative integers $\mathbb N$, phase vectors $e$ and $b_m=c_m/a_m$ (normalized and $m\in\overline{\mathcal N}$).
Each choice will deliver a particular solution.
This highly degenerate problem can be restricted to some principal solutions.

\subsection{Principal Solutions}
It is useful to restrict the manifold of solutions to the reasonable ones.
For this reason we consider properties of our operator $\rho$.
In particular, the operator $\rho$ is symmetric (exchange of the subsystems $A$ and $B$ delivers the same operator), and it consists only of real variables, $\rho_{m,n}=\rho_{m,n}^\ast$.
Thus, we may assume $|a,b\rangle=|a,a\rangle$, cf. a similar argumentation in~\cite{SymQuantumStates,SQS}, and only real vectors $e$.
This particular choice will be sufficient for the construction of the desired quasiprobability of entanglement.
Using the solutions given above we can now choose a set $\overline{\mathcal N}$, which yields
\begin{align}
	\nonumber&c=\rho_{\overline{\mathcal N}}^{-1}e
	\quad\Rightarrow\quad a_m=\pm_m\sqrt{c_m},\label{Eq:23}\\
	&|a\rangle=\sum_{m\in\overline{\mathcal N}} \frac{a_m}{\left[\sum_{m\in\overline{\mathcal N}}|a_m|^2\right]^{1/2}}|m\rangle,\\
	&g=\langle a,a|L|a,a\rangle,\nonumber
\end{align}
$\pm_m$ being the possible roots for each component, and $e=(e_n)_{n\in\overline{\mathcal N}}$ with
$e_n\in\{+1,-1\}$.
This form is useful for numerical purposes, since it only requires the inversion of the $N\times N$ matrix $\rho_{\overline{\mathcal N}}$, where $N=|\overline{\mathcal N}|$.

Let us comment on the trivial solutions.
Condition~(\ref{Eq:CondZero}) is fulfilled in the trivial case when for each $m$ holds either $a_m=0$, or $b_m=0$, or $a_m=b_m=0$.
From $a_mb_m=0$ follows that the SE value is $g=0$.
Note that these solutions include all SE values $g=0$, if we assume that the coefficient matrix $(\rho_{m,n})_{m,n\in\mathbb N}$ has full rank.

\subsection{Elementary Examples}

\paragraph*{Example $N=1$.}
For a better understanding of the solutions we may study the case $\overline{\mathcal N}=\{k\}$.
We insert this in the above general solution.
This leads to a SE vector $|a,a\rangle=|k,k\rangle$ in Fock basis and a SE value $g=\rho_{k,k}$.
However, we can directly check whether this is a solution.
The reduced operators used in the definition of the SE Eq.~(\ref{Eq:SEE}) are
\begin{align}
	\rho_a=\rho_b=\rho_{k,k}\,|k\rangle\langle k|.
\end{align}
Obviously, $|k\rangle$ solves the eigenvalue problem for each reduced operator.
It is noteworthy that these solutions for $\overline{\mathcal N}=\{k\}$ already include the maximal SE value.
This maximal SE value is needed to construct an optimized entanglement witness from the operator $\rho$, cf.~\cite{Spe2,Spe3}.

\paragraph*{Example $N=2$.}
A nontrivial example can be given for a set $\overline{\mathcal N}=\{k,l\}$.
This yields a matrix
\begin{align}
	\rho_{\{k,l\}}=&\left(\begin{array}{cc}
		\rho_{k,k}&\rho_{l,k}^\ast\\\rho_{l,k}&\rho_{l,l}
	\end{array}\right),\\
	\nonumber \Rightarrow
	\rho_{\{k,l\}}^{-1}=&\frac{1}{\det(\rho_{\{k,l\}})}\left(\begin{array}{cc}
		\rho_{l,l}&-\rho_{l,k}^\ast\\-\rho_{l,k}&\rho_{k,k}
	\end{array}\right).
\end{align}
Using the definition of $e=(e_k,e_l)^{\rm T}$, the SE vector $|a,a\rangle$ and SE value $g$ can be directly formulated as
\begin{align}
	\nonumber |a\rangle&=\frac{\sqrt{\rho_{l,l} e^{{\rm i}\phi_k}-\rho_{l,k}^\ast e^{{\rm i}\phi_l}}|k\rangle\pm\sqrt{-\rho_{l,k} e^{{\rm i}\phi_k}+\rho_{k,k} e^{{\rm i}\phi_l}}|l\rangle}{\sqrt{|\rho_{l,l} e^{{\rm i}\phi_k}-\rho_{l,k}^\ast e^{{\rm i}\phi_l}|+|-\rho_{l,k} e^{{\rm i}\phi_k}+\rho_{k,k} e^{{\rm i}\phi_l}|}},\\
	g&=\frac{\rho_{k,k}+\rho_{l,k}e^{{\rm i}(\phi_k-\phi_l)}+\rho_{l,k}^\ast e^{-{\rm i}(\phi_k-\phi_l)}+\rho_{l,l}}{|\rho_{l,l} e^{{\rm i}\phi_k}-\rho_{l,k}^\ast e^{{\rm i}\phi_l}|+|-\rho_{l,k} e^{{\rm i}\phi_k}+\rho_{k,k} e^{{\rm i}\phi_l}|},
\end{align}
with using the fact that a global phase can be ignored.
Note that a global phase transformation can be performed for the state $|a\rangle$.
The solution in this case only depends on the phase difference $\phi_k-\phi_l$.

\section{Application to dephased two-mode squeezed-vacuum states}\label{Sec:Visu}
In this section we apply our method to the dephased two-mode squeezed vacuum state $\varrho_{\sigma}$ given in Eq.~(\ref{eq:dephased-sv}).
For the identification of entanglement of a general CV entangled state it has been proven that it is necessary and sufficient to identify entanglement in finite subspaces of the Hilbert space of the system under study, for details see~\cite{fin-sp}.
This is very helpful in practice.
Making use of this result, entanglement of a non-Gaussian bipartite CV quantum state could be already demonstrated in two-qubit subspaces~\cite{Schnabel3}.

Based on this knowledge, we start to consider the entanglement quasiprobabilities in the Fock spaces containing zero and one photon per mode.
This subspace has the highest probability to be measured, and the interference terms are the maximal ones, cf. Eq.~(\ref{eq:dephased-sv}).
In Fig.~\ref{Fig:Dephasing} we illustrate the phase distributions chosen for our examples.
They are ranging from a $\delta$~distribution for $\sigma=0$ towards an almost uniform phase distribution for $\sigma=5$.
\begin{widetext}
\begin{center}
	\begin{figure}[ht]
		\includegraphics[width=4.2cm]{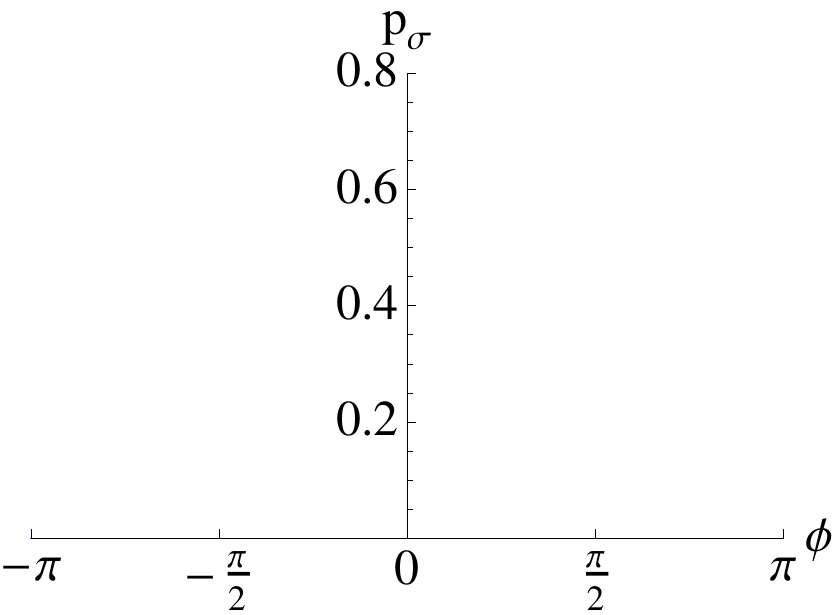}\hspace*{0.2cm}
		\includegraphics[width=4.2cm]{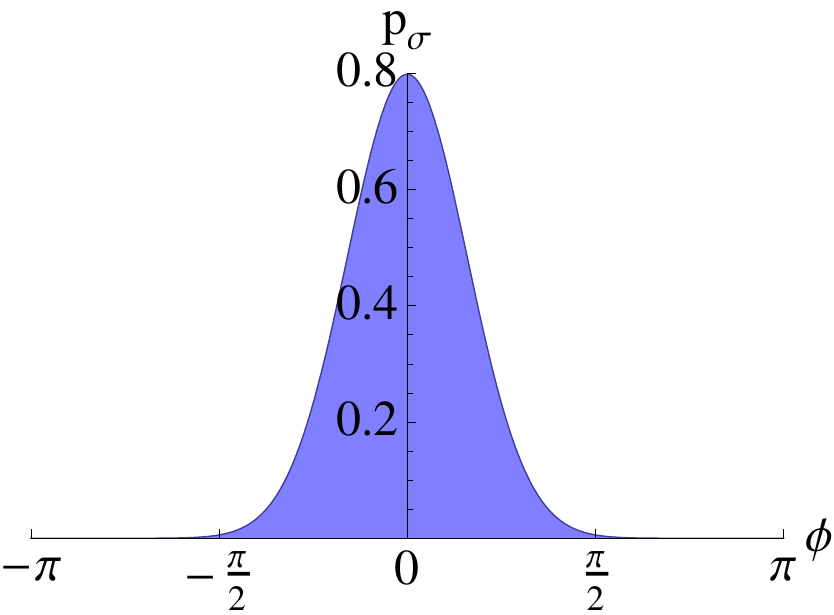}\hspace*{0.2cm}
		\includegraphics[width=4.2cm]{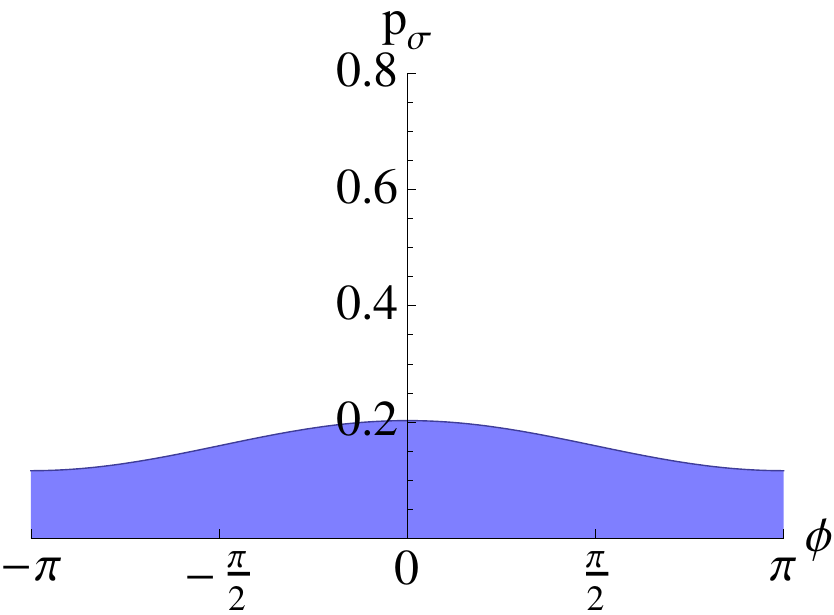}\hspace*{0.2cm}
		\includegraphics[width=4.2cm]{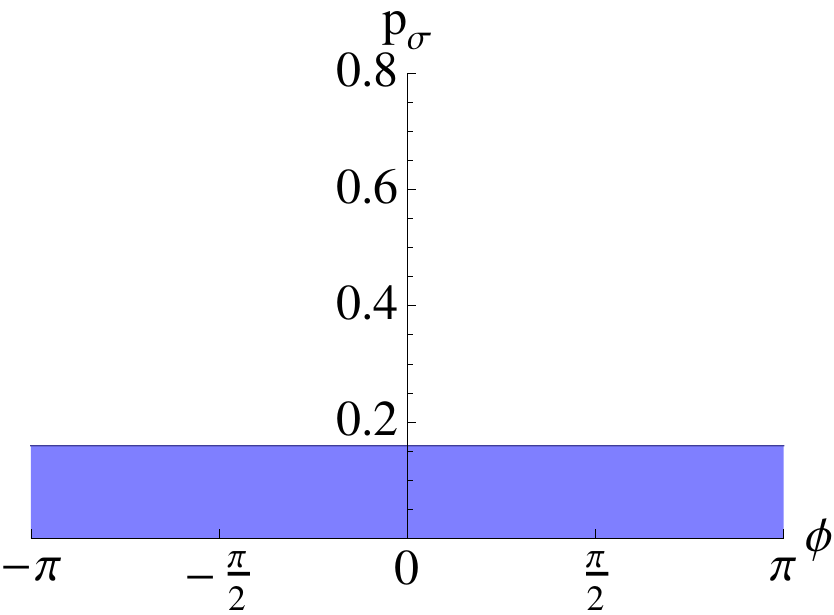}
		\caption{
		The distributions $p_\sigma(\phi)$ are shown for the considered phase randomization, cf. Eq.~(\ref{Eq:PeriodicGaussian}), for $\sigma=0,0.5,2,5$ (from left to right).
		The first plot represents a $\delta$ distribution centered at the origin.
		The last plot is close to a full randomization with an almost uniform phase distribution.
		}\label{Fig:Dephasing}
	\end{figure}
\end{center}
\end{widetext}

In Fig.~\ref{Fig:2x2} we show the entanglement quasiprobabilities for the two-qubit spaces under consideration.
The first three distributions show clear negativities, which tend to decrease with increasing dephasing.
These negativities uncover the entanglement of our dephased squeezed vacuum state $\varrho_\sigma$, not only for the chosen subspace but even for the complete CV state.
For a full phase randomization the negativities disappear, since the corresponding state is separable.
The plot for $\sigma=5$ is very close to this situation -- the negativities are too tiny to be visible.
\begin{widetext}
\begin{center}
	\begin{figure}[ht]
		\includegraphics[width=4.2cm]{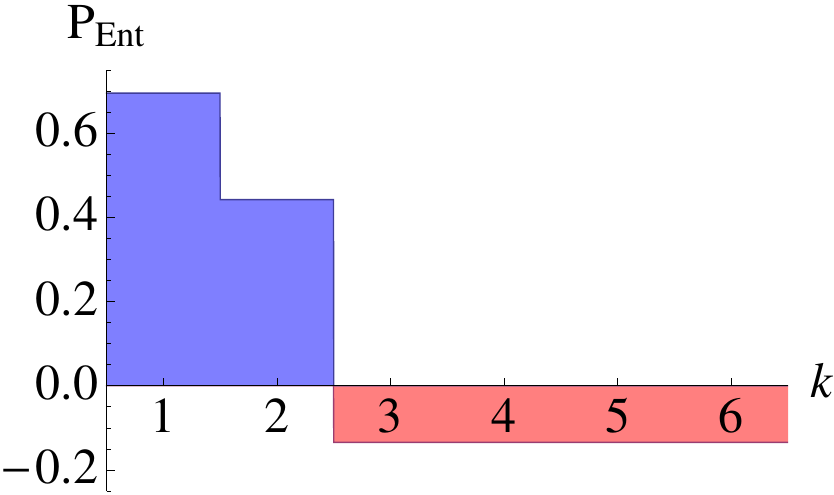}\hspace*{0.2cm}
		\includegraphics[width=4.2cm]{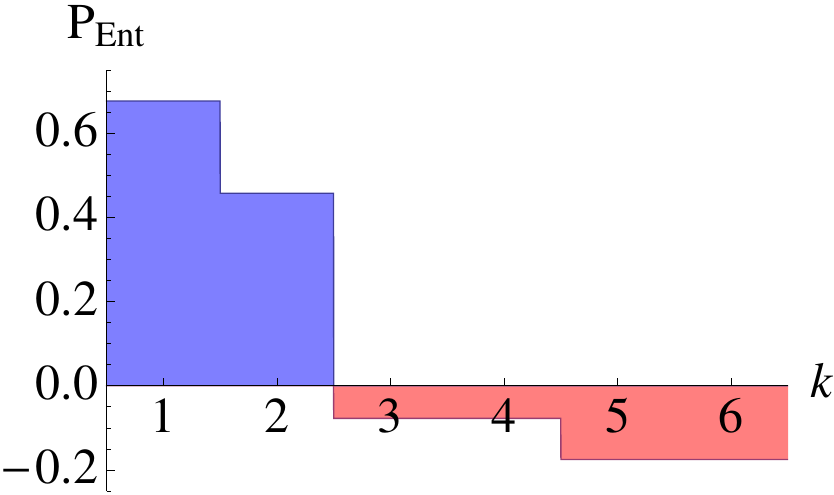}\hspace*{0.2cm}
		\includegraphics[width=4.2cm]{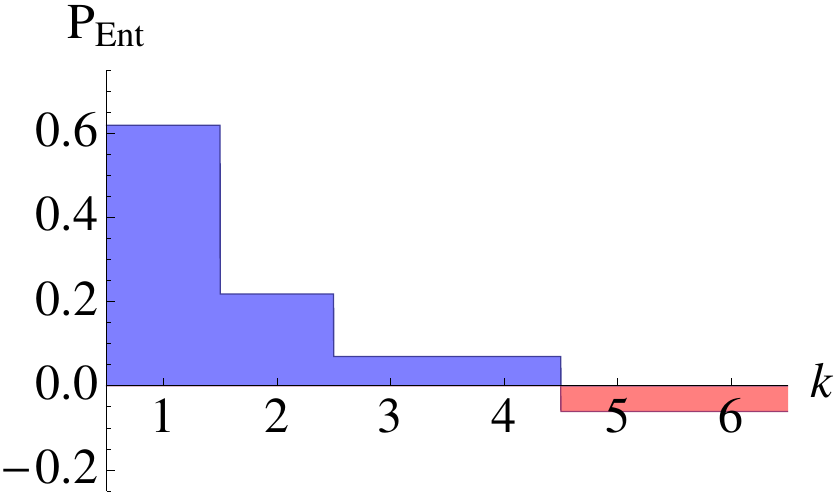}\hspace*{0.2cm}
		\includegraphics[width=4.2cm]{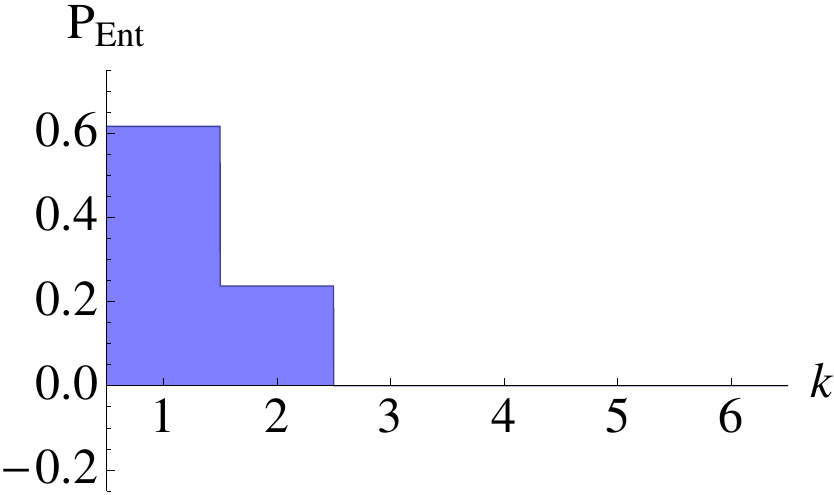}
		\caption{
		The plot shows the optimized entanglement quasiprobabilities for different values $\sigma=0,0.5,2,5$ of dephasing, for a squeezed vacuum state with $\zeta=0.62$, corresponding to $-6.3\,{\rm dB}$ noise reduction.
                The ordering is the same as in Fig.~\ref{Fig:Dephasing}.
		The values of $k$ number the different solutions of the SE equations.
		The corresponding sets $\overline{\mathcal N}_k$ are: $\overline{\mathcal N}_1=\{0\}$, $\overline{\mathcal N}_2=\{1\}$, $\overline{\mathcal N}_{3,4,5,6}=\{0,1\}$.
		For the latter set the individual solutions are discriminated by different choices of the signs $\pm$.
		}\label{Fig:2x2}
	\end{figure}
\end{center}
\end{widetext}

In Table~\ref{Tab:EXample} we give the SE values, the coefficients of the SE states, and the quasiprobabilities, for the case of dephasing with $\sigma=2$.
The numerical values are obtained from Eq.~(\ref{Eq:23}).
Taking into account that a global phase can be ignored, we obtain the given six SE vectors.
Comparing the reconstructed state $\varrho_{{\rm rec}, \sigma}$ according to Eq.~(\ref{Eq:DefDecPEnt}) with the original, but truncated state $\varrho_\sigma$ as given in Eq.~(\ref{eq:dephased-sv}) yields a numerical error of $\epsilon=4\times10^{-16}$ (precision of double-floating-point numbers).
The numerical error is computed in the Hilbert-Schmidt norm $\epsilon=\left[{\rm Tr}(\varrho_{{\rm rec}, \sigma}-\varrho_{\sigma})^2/{\rm Tr}\varrho_{\sigma}^2\right]^{1/2}$.
\begin{widetext}
\begin{center}
	\begin{table}[ht]
	\begin{tabular}{|l|cccccc|}
		\hline
		$k$ & 1&2&3&4&5&6\\
		\hline\hline
		SE value  $g_k$ & 0.615600 & 0.236637 & 0.190946 & 0.190946 & 0.149659 & 0.149659 \\
		\hline
		SE vector $|a_k\rangle$ & 1.000000 & 0.000000 & 0.496987 & 0.496987 & 0.549275 & 0.549275 \\
		& 0.000000 & 1.000000 & 0.867758 & -0.867758 & 0.835642i & -0.835642i \\
		\hline
		$P_{\rm Ent}(a_k)$ &0.618287 & 0.217677 & 0.069431 & 0.069431 & -0.061295 & -0.061295\\
		\hline
	\end{tabular}
	\caption{
		The table contains the explicit numerical solution for a phase randomization $\sigma=2$.
		The columns include the information for the $k$-th solution.
		The two components $a_{k,0}$ and $a_{k,1}$ of the SE vector $|a_k\rangle=a_{k,0}|0\rangle+a_{k,1}|1\rangle$ are given in the third and fourth line, respectively.
	}\label{Tab:EXample}
	\end{table}
\end{center}
\end{widetext}

\begin{center}
	\begin{figure}[ht]
		\includegraphics[width=8.4cm]{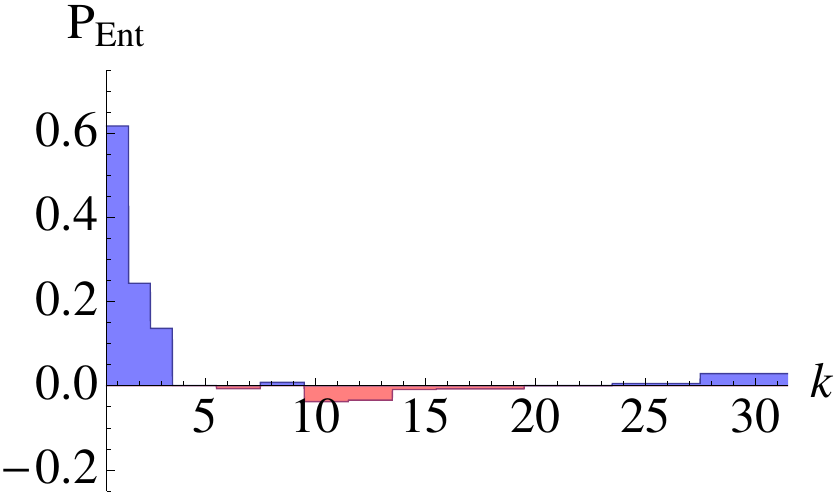}
		\caption{
		The quasiprobability of $\varrho_{\sigma}$ is shown for $\sigma=5$ and a squeezing parameter $\zeta=0.62$.
		The values of $k$ number the different solutions of the SE equations, according to the  sets: $\overline{\mathcal N}_1=\{0\}$, $\overline{\mathcal N}_2=\{1\}$, $\overline{\mathcal N}_3=\{2\}$, $\overline{\mathcal N}_{4,\dots,7}=\{0,1\}$, $\overline{\mathcal N}_{8,\dots,11}=\{0,2\}$, $\overline{\mathcal N}_{12,\dots,15}=\{1,2\}$, $\overline{\mathcal N}_{15,\dots,31}=\{0,1,2\}$.
		}\label{Fig:Delta5}
	\end{figure}
\end{center}

In cases when the entanglement quasiprobabilities of the two-qubit subspaces do not show significant negativities, we can proceed to study subspaces of higher dimensions.
Let us consider two-qutrit subspaces including the photon numbers of up to two per radiation-field mode.
In Fig.~\ref{Fig:Delta5} we deal with an example for relatively strong dephasing, $\sigma=5$.
The entanglement quasiprobability in the two-qubit subspace fails to show significant negativities, see Fig.~\ref{Fig:2x2}.
Our calculation yields a maximal negativity of $-1.773\times10^{-6}$.
However, in the two-qutrit ($3\otimes3$) subspace the entanglement quasiprobability becomes clearly negative, with a maximum negativity of $-3.786\times10^{-2}$.
Surprisingly, even for such strong dephasing our method may identify some residual entanglement in the dephased two-mode squeezed vacuum state under study.
In principle our method can be further extended to higher dimensional subspaces whenever needed to increase the significance of the negativities, at the expense of increasing numerical complexity.

Let us conclude this section with a general remark concerning the use of entanglement quasiprobabilities.
We have outlined in Sec.~\ref{Sec:EntQP} that the entanglement quasiprobabilities yield a complete representation of a general bipartite quantum state, cf.~Eq.~(\ref{eq:ent-repr}).
In the successive treatment of subsystems of increasing dimensions, however, of course we only get a complete representation of the reduced quantum state in the chosen subspace.
Another important issue is the quantification of entanglement.
In a very general sense entanglement can be quantified by the Schmidt number (SN), cf.~\cite{Terhal,Spe4,Shirokov}.
A method for determining the SN has been introduced recently~\cite{Spe3}.
In fact, the quasiprobability method developed here can be extended to identify the SN via negativities of SN-quasiprobabilities.

\section{Summary and conclusions}\label{Sec:SandC}
We have derived the entanglement quasiprobabilities for a nontrivial entangled quantum state.
This distribution visualizes nonclassical correlations between radiation modes.
Our studied state consists of a two-mode squeezed vacuum undergoing dephasing in one channel.
The resulting state is non-Gaussian, which is of some interest for various applications in quantum technology.
We provide a method for solving the separability eigenvalue problem of the quantum state under study.
From the solution we could derive the entanglement quasiprobabilities of the state.
Their negativities are necessary and sufficient for entanglement of the truncated quantum state in a chosen subspace.
Moreover, the existence of negativities in some subspace is necessary and sufficient for the entanglement of the infinite dimensional continuous-variable quantum state.

We have identified entanglement of dephased squeezed states in two-qubit subspaces.
For strong dephasing, however, the corresponding entanglement quasiprobabilities may only show tiny negativities which are hard to determine in experiments.
In such a case we have shown that the entanglement quasiprobabilities in a two-qutrit subspace may show negativities which are enhanced by a factor of about $2\times 10^4$ compared with the two-qubit space.
Beside the more significant entanglement signatures, the extension of our method to higher dimensions is of fundamental interest.
For this purpose, we formulated the general procedure and the solution of the separability eigenvalue equations.
It allows one to identify quantum correlations for the important class of radiation-field states under study, in the notion of bipartite entanglement in its most general form.

\section*{Acknowledgments}
\noindent The authors gratefully acknowledge financial support by the Deutsche Forschungsgemeinschaft through SFB 652.

\end{document}